\documentclass[12pt]{article}
\input epsf
\newcommand{\beq}{\begin{eqnarray}}
\newcommand{\eeq}{\end{eqnarray}}

\newcommand{\bi}{\bibitem}

\newcommand{\be}{\begin{equation}}
\newcommand{\ee}{\end{equation}}
\newcommand{\ben}{\begin{eqnarray}\displaystyle}
\newcommand{\een}{\end{eqnarray}}

\def\sqr#1#2{{\vcenter{\vbox{\hrule height.#2pt
         \hbox{\vrule width.#2pt height#1pt \kern#1pt
            \vrule width.#2pt}
         \hrule height.#2pt}}}}

\begin{document}

{}~ \hfill\vbox{\hbox{hep-ph/0409133} \hbox{PUPT-2135} }\break

\vskip 0.5cm

\begin{center}
\Large{\bf Superconformal Gauge Theories and Non-Critical Superstrings}

\vspace{10mm}

\normalsize{Igor R. Klebanov}

\vspace{2mm}

\normalsize{\em Joseph Henry Laboratories, Princeton University}

\vspace{0.2cm}
\normalsize{\em Princeton, NJ 08544}

\vspace{7mm}

\normalsize{Juan M. Maldacena}

\vspace{2mm}
\normalsize{\em Institute for Advanced Study}

\vspace{0.2cm}

\normalsize{\em Princeton, NJ 08540}
\end{center}

\vspace{10mm}

\begin{abstract}
\medskip
We consider effective actions for six-dimensional non-critical
superstrings. We show that the addition of $N$ units of R-R flux
and of $N_f$ space-time filling D5-branes produces $AdS_5 \times S^1$
solutions with curvature comparable to the string scale.
These solutions have the right structure to be dual to ${\cal N}=1$
supersymmetric $SU(N)$ gauge theories with $N_f$ flavors. We
further suggest bounds on the mass-squared of tachyonic fields in
this background that should restrict the theory to the conformal window.

\end{abstract}

\newpage

\section{Introduction}

Most known examples of the AdS$_5$/CFT$_4$ correspondence
\cite{JM,GKP,EW} relate conformal 4-d gauge theories to
$AdS_5\times X^5$ backgrounds of the {\it critical} type IIB
strings with R-R flux. This correspondence may be motivated by
considering stacks of D3-branes and taking the low-energy limit
\cite{reviews}. When the `t Hooft coupling of the gauge theory is
made large, the curvature becomes small in the dual description,
so that many calculations can be performed using the supergravity
approximation to the type IIB string theory. This duality provides
a great deal of information about strongly coupled gauge theory.
However, the simplest methods apply only to theories that can be
taken to very strong `t Hooft coupling. One expects that there are
many superconformal gauge theories that do not satisfy this rather
stringent requirement. In fact, many of the ${\cal N}=1$
superconformal gauge theories discovered by Seiberg \cite{Seiberg}
do not have known AdS duals.

For this reason it is interesting to search for string theories other
than the conventional critical ones, as starting points for the AdS/CFT
duality. In this connection Polyakov proposed to use non-critical
type 0 string theory \cite{Polyakov}. By considering the effective action
of this theory, he provided evidence for the existence of an $AdS_5$ solution
whose dual should be a non-supersymmetric conformal gauge theory.
Furthermore, there exist non-critical string
solutions of the form $AdS_p\times S^k$, $k>1$,
supported by R-R flux \cite{Polyakov,Kuperstein}, and $AdS_p$ times
tori \cite{Silverstein}.\footnote{
Also, backgrounds of the form
$AdS_p$ times linear dilaton were recently found in \cite{Janus}.}
The curvature of the AdS space, however,
turns out to be of order the string scale, so there
may be large corrections to the 2-derivative effective action approximation.
Nevertheless, the effective action may still lead to interesting qualitative
information about the dual gauge theory.

In this note we consider a generalization of the effective action
approach to ${\cal N}=1$ superconformal gauge theories. Since such
theories have a $U(1)$ R-symmetry, the simplest geometry they may
be dual to is $AdS_5\times S^1$, with the R-symmetry realized by
translations in $S^1$.   A sigma model construction of
this background, as well as of AdS backgrounds
in other dimensions, was recently presented in
\cite{Polyakovnew}. In search of $ AdS_5 \times S^1 $
 geometries, we consider
backgrounds created by D-branes in the 6-dimensional non-critical
superstring theory. Non-critical superstrings were studied in
\cite{Kut,Ooguri,Giveon,Hori} and recently reviewed in
\cite{Murthy}. The 6-dimensional case is of particular interest.
In the absence of D-branes, it is believed \cite{Ooguri,Giveon}
that its background $R^{3,1}\times SL(2,R)/U(1)$ is dual to type
IIB string theory on the resolved conifold; the mirror background
$R^{3,1}$ times ${\cal N}=2$ super-Liouville theory is dual to IIB
string theory on the deformed conifold.

Starting with the type IIB theory, the ${\cal N}=1$ supersymmetric
$SU(N)$ gauge theory may be realized on $N$ D5-branes wrapped over
the 2-cycle of the resolved conifold. The backgrounds they create
were studied in \cite{KS,MN,Vafa} where it was was demonstrated
that the theory makes a geometric transition to the deformed
conifold. Furthermore, by adding D5 \cite{CIV,Cachazo,nunez} or D7
branes \cite{Grana,Karch,Kruczenski,Sakai,Ouyang,Burrington}, it
is possible to introduce flavors into the gauge theory. In this
paper we use similar methods to embed ${\cal N}=1$ SYM theory into
the 6-dimensional non-critical superstring. The theory without
matter may be realized on a stack of $N$ D3-branes of the 6-d
superstring placed at the tip of the $SL(2,R)/U(1)$ cigar. The
flux produced by these D3-branes warps and deforms the
$R^{3,1}\times SL(2,R)/U(1)$ geometry; we study its effect using
the minimal 2-derivative effective action of the 6-d non-critical
superstring. We do not find an $AdS_5 \times S^1$ solution; this
could be expected since there is no candidate superconformal gauge
theory without flavors. This motivates us to add $N_f$ space-time
filling uncharged D5 branes which realize $N_f$ flavors in the SYM
theory. This system possesses $AdS_5 \times S^1$ solutions for all
values of ${N_f\over N}>0$. We further suggest that requiring the
mass-squared of open string tachyons to lie in the range
$-4 < (m R_{AdS})^2< -3$,
where both the $\Delta_+$ and the $\Delta_-$ quantizations
are admissible \cite{BF,KW}, restricts these backgrounds
to the Seiberg conformal window \cite{Seiberg}.

A circumstance that complicates our analysis is that, as in \cite{Polyakov},
the curvature of $AdS_5$
is of order the string scale (with the 2-derivative
effective action, we find that $R^2_{AdS}=6 \alpha'$
for all $N$ and $N_f$). Therefore, our results are expected to receive
$O(1)$ corrections from higher-derivative terms. Nevertheless, it is plausible
that such corrections do not destroy our $AdS_5\times S^1$ solutions.
%and that the connection between the stability in $AdS_5$
%and the existence of a conformal window in the SYM theory
%continues to hold.

\section{Effective Action of the Non-Critical String}

Let us consider the following effective action for the 6-d
non-critical superstring theory,
\be  \label{effa}
S\sim \int d^6 x \sqrt {-G} [ -
(\partial_\mu \chi)^2+ e^{-2\phi} (R + 4 (\partial_\mu \phi)^2 +
\lambda^2 ) - 2 N_f e^{-\phi}] \ , \ee where $\chi$ is a R-R
scalar dual to the 5-form field strength.\footnote{ We did not compute
the precise numerical coefficients in the normalization of the
RR terms and of the D5-brane tension.} The cosmological constant is
$$\lambda^2 = {10-d\over \alpha'} = 4
\ ,
$$
in units where $\alpha'=1$.
We will assume that $\chi = N\theta$.
This means that we have $O(N)$ units of RR flux, so
the number of colors in the gauge theory is
proportional to $N$.
The last term in (\ref{effa})
is due to uncharged D5-branes, which can be thought of
as D5/anti-D5 pairs,
filling all six space-time dimensions.
So, the parameter $N_f$ is proportional to the number
of massless flavors in the gauge theory.

Let us transform to the Einstein metric,
\be
 g_{\mu \nu} = e^{-\phi} G_{\mu \nu}
\ .
\ee
The action becomes
\be
\int d^6 x \sqrt {-g} [ - e^{2\phi} (\partial \chi)^2 +
R - (\partial \phi)^2 + e^{\phi} \lambda^2 -
2 N_f e^{2\phi} ]
\ .
\ee
We adopt the following ansatz for the 6-d Einstein metric:
\be
 ds^2_{E} = \bigg[  e^{-2B(u)/3}
  (du^2 + e^{2 A(u)} dx_i dx_i) + e^{2B(u)} d\theta^2 \bigg]\ ,
\ee
Substituting this into the action, we find
\be S \sim  \int du e^{4A} [ 3 (A')^2- {1\over 3} (B')^2- {1\over
4}(\phi')^2 + {1\over 4}\lambda^2 e^{\phi-(2B/3)}- {1\over 4} N^2
e^{2\phi-(8B/3)} -{N_f\over 2} e^{2\phi - (2B/3)} ] \ . \ee
When the action takes the  form
\be \label{genact}
S =  \int du \ e^{4
A} \bigg[ 3 A'^2 - { 1 \over 2} {\bf G}_{ab}(f)  f'^a  f'^b -
 V(f)\bigg]
 \ ,
\ee
there is the following prescription for looking for
first-order equations \cite{Freedman,Skenderis,DeWolfe}.
If we find the superpotential $W$, then
\be
f'^a = { 1 \over 2} {\bf G}^{ab} { \partial  W \over \partial f^b} \ ,
\ \ \ \ \ \ \ \
A' = - { 1 \over 3} W (f) \ ,
\ee
where the superpotential $W$ is a
function of scalars $f^a$
satisfying
\be \label{genrel}
V = {1\over 8} {\bf G}^{ab} {\partial W\over \partial f^a}
{\partial W\over \partial f^b} - {1\over 3} W^2
\ .
\ee
Here we have only two scalars: $B$ and $\phi$.
\be \label{fieldmet}
{\bf G}_{\phi \phi} = {1 \over 2}\ ,
{\bf G}_{BB} = {2 \over 3} \ .
\ee
The potential is
\be \label{genpot}
V =
-{\lambda^2\over 4}
e^{\phi-(2B/3)}+{N^2\over 4} e^{2\phi-(8B/3)}
+{N_f\over 2} e^{2\phi - (2B/3)}
\ .
\ee

If $N=N_f=0$ then the superpotential may be taken of the form
\be W\sim \lambda\exp[(\phi/2) - (B/3)]
\ .
\ee
This gives a system of first-order equations.
The resulting solution is that the string frame metric is exactly
flat and the dilaton linear
\beq ds^2 &=& dx_i dx_i +  d \rho^2  + d\theta^2  \ ,
\\
\phi &=& - \rho \ ,
\eeq
where $\rho = { 3 \over \lambda} \log u$.

To get the well-known ``cigar'' solution \cite{Elitzur} times $R^4$,
we need to adopt a different
superpotential:
\be W = - \exp[-4B/3] - \exp[\phi + (2B/3)]
\ ,
\ee
which also gives $V=-\exp[\phi-(2B/3)]$.
We define a new variable $g$ and  a new radial variable $r$ through
\beq dr &=&  \label{rdef}
 \exp[(\phi/2)-(B/3)] du\ ,
\\
g &=& {\phi\over 2} + B \ .\nonumber
\eeq
We see that $e^{g}$ is the radius of the circle in string units.
We find the following set of first order equations:
\beq
 {\partial\phi\over \partial r} &=&- e^g\ ,
\\
 {\partial g\over \partial r} &=& - e^g + e^{-g} \ .
\eeq
They have a solution
\be e^{- \phi} =\cosh r\ ,\ \ \ \ \ ~~~~ e^g = \tanh r \ .
\ee
So the string frame metric is
\be ds^2 = dx_i dx_i + dr^2 + \tanh^2 r d\theta^2\ ,
\ee
which  is precisely the 2-d black hole metric.

\section{The $AdS_5\times S^1$ Solutions}

For general $N$ and $N_f$ there exists an $AdS_5\times S^1$ solution
where the dilaton and the radius of $S^1$ have $u$-independent
values
which extremize the effective potential:
\be {\partial V\over \partial \phi}= {\partial V\over \partial B}= 0
\ .
\ee
These equations are
\be \label{exone} {2\over 3}
e^{\phi-(2B/3)}-{2 N^2\over 3} e^{2\phi-(8B/3)}
-{N_f\over 3} e^{2\phi - (2B/3)} =0\ ,
\ee
\be \label{extwo}
-e^{\phi-(2B/3)}+{ N^2\over 2} e^{2\phi-(8B/3)}
+N_f e^{2\phi - (2B/3)} =0\ .
\ee
The solution is
\be \label{adssol}
e^{2 B} = {N^2\over N_f}\ , \qquad
e^\phi= {2\over 3 N_f}
\ .
\ee
Now, we define the variables $r$ and  $g$  as above, (\ref{rdef}), and
\begin{equation}
f = {\phi\over 2} - {B\over 3} + A
\ .\end{equation}
In terms of these new variables the ansatz for the string metric
is
\begin{equation}
 ds^2 = e^{2 f} dx_i dx_i + dr^2 + e^{2 g} d\theta^2
\ .\end{equation}
From the constraint associated with the action (\ref{genact}),
\be 3 \left
({\partial A\over \partial u}\right )^2 -
{ 1 \over 2} {\bf G}_{ab}(f)
{\partial f^a\over \partial u}
{\partial f^b\over \partial u} + V(f)=0\ ,
\ee
we find
\be
\left ({\partial f\over \partial r}\right )^2 = {1\over 6}
\ .
\ee
This implies that in the string frame the AdS radius is
$\sqrt{ 6\alpha'}$.
As $N_f \to 0$ this solution becomes singular. Indeed,
for $N_F=0$ and $N>0$ there is no $AdS_5\times S^1$ solution
of the equations, since (\ref{exone}) and (\ref{extwo}) are
incompatible.

The solution above could also be found directly from the Einstein equations.
Starting with the Einstein frame, we find that the dilaton
variational equation is
\be 2 e^{2\phi} [ 2N_f + (\partial \chi)^2] = 4 e^\phi
\ ,
\ee
while the Einstein equation is
\be R_{\mu\nu}= {1\over 2} g_{\mu\nu}
(R- e^{2\phi} (\partial \chi)^2 - 2 N_f e^{2\phi} + 4 e^{\phi})
+ e^{2\phi} \partial_\mu \chi \partial_\nu \chi
\ .
\ee
It follows that
\be
R= e^{2\phi} (\partial \chi)^2 + 3( N_f e^{2\phi} - 2 e^{\phi})
\ee
and
\be R_{\mu\nu}= {1\over 2} g_{\mu\nu}
( N_f e^{2\phi} - 2 e^{\phi})
+ e^{2\phi} \partial_\mu \chi \partial_\nu \chi
\ .
\ee
Now it is easy to check that the RHS of the equation for
$R_{\theta\theta}$ is indeed zero when we substitute
$\chi= N \theta$ and
the solution (\ref{adssol}) for $B$ and $\phi$.
The dilaton equation is also satisfied.

The equation for the remaining components $R_{ij}$
indicate that it is a space of constant negative curvature:
\be
R_{ij}=-{4\over 9 N_f} g_{ij}
\ .
\ee
After we transform back to the string frame, we have
\be
R_{ij}=-{4\over 6 } G_{ij}=- {4\over R_{AdS}^2} G_{ij}
\ .
\ee
Thus, all dependence of the $AdS_5$ radius
on $N_f$ and $N$ cancels out,
and we again find that it is $\sqrt{6\alpha'}$.\footnote{
Note that this result does not depend on the numerical coefficients
in the RR flux quantizations, which we have not fixed.}
In summary, for this solution the string frame radii of the
$S^1$ and the $AdS_5$ are
\be
{ R^2_{S^1} \over \alpha'} =  e^{2 g } = { 2 \over 3} { N^2 \over N_f^2}
\ .\ee
\be
{ R^2_{AdS} \over \alpha'}  = 6\ .
\ee
The 5-dimensional effective coupling is determined by
\be
{1\over g_5^2} = e^{-2\phi} R_{S^1}=
{3\sqrt{ 3\alpha'}\over 2\sqrt 2} N N_f\ .
\ee

It is of further interest to
calculate the masses of the fluctuations in $\phi$ and $B$
around the minimum (\ref{adssol}) of the potential. We
find that $N_f$ and $N$ scale out of the calculation,
so that the masses are completely independent of $N_f$ and $N$:
\be  m_1^2 = {5+\sqrt {13}\over 3 \alpha'}\ , \qquad
m_2^2 = {5-\sqrt {13}\over 3 \alpha'}\ .
\ee
The corresponding fields are certain mixtures of $\phi$ and $B$.
Since both $m^2$ are positive, the corresponding operator dimensions
are determined by the positive branch, $\Delta_+= 2+\sqrt{4+ (m R_{AdS})^2}$,
and we find
\be \label{opdim}  \Delta_1 = 3+\sqrt {13}\ , \qquad
\Delta_2 = 1+\sqrt {13}\ .
\ee
The dimensions are not rational, which
is likely due to the fact that our solution undergoes $O(1)$ corrections
by higher-derivative terms.

\section{First Order Equations }

There exists a simple first order formulation of the
equations we need to solve.
We consider the following superpotential:\footnote{In the original version of this paper
the superpotential was written down for ${N_f\over N}=1$. We thank Stanislav Kuperstein
and Cobi Sonnenschein for sending us its generalization to arbitrary ${N_f\over N}$.}
\be \label{newsup}
W = - {N\over N_f} \exp[-4B/3] - {N_f\over N} \exp[\phi + (2B/3)]
+  N \exp[\phi - (4B/3)]
\ .
\ee
Using (\ref{genrel}) and (\ref{fieldmet}) we get the potential
(\ref{genpot}).

From the superpotential (\ref{newsup}) we get the first-order equations
\begin{equation}
 {\partial g\over \partial r} = {N\over N_f} e^{ -g } - {N_f\over N} e^g -{1\over 2} N e^{\phi-g}\ ,\qquad
{\partial \phi\over \partial r} =  - {N_f\over N} e^{g} +  N e^{\phi - g}\ , \qquad
{\partial f\over \partial r} =  {1\over 2} N e^{ \phi - g}
\  . \label{susyeqn}
\end{equation}
We note that there exists an $AdS_5 \times S^1$ solution with
$e^{2g_c}=  { 2 N^2 \over 3 N_f^2}$ and
$ N_f e^{\phi_c} = { 2 \over 3} $.
This means that $e^{2 B_c}= {N^2\over N_f}$.

It is possible to study the flows associated to the equation
(\ref{susyeqn}). We were not able to find them analytically.
As a first step we can perturb the equation around the
point at $(g_c,\phi_c)$. Deviations around this AdS fixed point are governed
by a linear equation for $ (\delta g , \delta \phi) $ given in terms of a matrix
\be
M = \sqrt{ 2 \over 3} \left( \begin{array}{ll} - 2 & - {1 \over 2} \\
 -2  & 1 \end{array} \right)
\ee
with eigenvalues
\be
\lambda_\pm = { 1 \over \sqrt{6}} \left( -1  \pm \sqrt{ 13 } \right)
\ee
We see that $\lambda_+$ is positive but $\lambda_-$ is negative.
Defining the standard radial variable $z=e^{-r/\sqrt 6}$ such that the AdS
metric is $6 (dz^2+ dx^2)/z^2$, we find that one of the perturbations grows in
the UV as $z^{-\sqrt 6 \lambda_+}$. Comparing with the operator dimensions
(\ref{opdim}) we note that this behavior is $z^{4-\Delta_1}$ corresponding
to adding an irrelevant operator to the action.
The other perturbation instead grows in the IR as
$z^{-\sqrt 6 \lambda_-}$. This behavior is $z^{\Delta_2}$ corresponding to
giving an expectation value to an operator of dimension $\Delta_2$ \cite{KW}.

If we integrate the flow starting from the AdS region at
very negative values of $r$, then we need to
move away from the IR fixed point in the direction corresponding to the eigenvector
of $M$ with positive eigenvalue, i.e. we forbid the perturbation that grows
in the IR as $z^{\Delta_2}$.
Integrating the equations numerically we find
that this leads to the cigar solution at large values of $r$. We see a plot
of the resulting function in figure 1.
In conclusion, there is a unique flow between the linear dilaton
region and the AdS region.

\begin{figure}[htb]
\begin{center}
\epsfxsize=3.0in\leavevmode\epsfbox{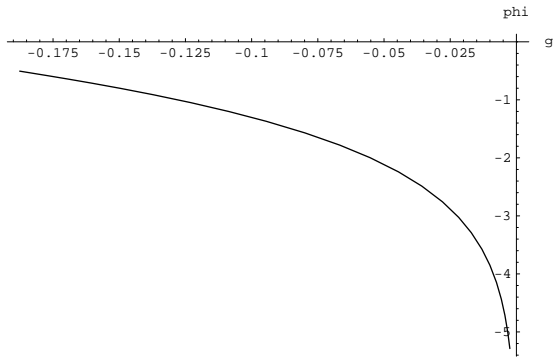}
\end{center}
\caption{  Flow between the $AdS_5 \times S^1$ and $R^4 \times $(linear dilaton).
The horizontal axis is $g$ and the vertical axis is $\phi$. In the
upper left-hand side the flow reaches the $AdS$ fixed point;
in the low right-hand side it asymptotes to the vertical
line. Thus, in the UV $g\rightarrow 0$ and the
coupling becomes very weak.  } \label{flows}
\end{figure}

\section{Stability, Unitarity and the Conformal Window}

The stability of our $AdS_5\times S^1$
solutions is an important issue since the theory
contains fields with $m^2 < 0$, and it is important to enforce
the Breitenlohner-Freedman \cite{BF} condition $m^2 R_{AdS}^2 \geq -4$. Indeed,
the theory on the cigar with the appropriate GSO projections
(see \cite{Murthy} for a discussion) contains the ordinary tachyon with
one unit of momentum or one unit of winding on the cigar.
Our first task is to calculate the effective mass of these modes
in our solutions.

Let us first assume the minimal string frame action for the
tachyon:
\be
-{1\over 2}\int d^6 x \sqrt{-G} [e^{-2 \phi} ( (\partial_\mu T)^2 +
m_0^2 T^2 ) ]
\ ,
\ee
where $m_0^2 =-2$ (in units where $\alpha'=1$).
The masses of the tachyons with one unit of momentum and one unit
of winding are:
\be
m_{1,0}^2 =  -2 + { 1 \over R^2_{S^1} } ~,~~~~~~~~
m_{0,1}^2 = -2 + { R_{S^1}^2 }
\ee
Since $R^2_{AdS}=6$,
we see that for any $R_{S^1}$ the winding and/or the momentum
mode violates the B-F bound. This seems to lead to the unfortunate
conclusion that the solutions we found are unstable.

We believe that the solution of this problem is related to non-minimal
couplings of the tachyon field. Indeed, additional terms coupling
the tachyon to R-R fields were found in \cite{KT}.
In the 6-d string case the extra term must have the form
\be
-{1\over 2}\int d^6 x \sqrt{-G}
b |F_5|^2 T^2
\ ,
\ee
where $b$ is a normalization constant.
Dualizing $F_5$ to the R-R scalar,
we see that the action becomes
\be \label{rrc}
-{1\over 2}\int d^6 x \sqrt{-G} [e^{-2 \phi} ( (\partial_\mu T)^2 +
m_0^2 T^2 ) + b (\partial_\mu \chi)^2 T^2 ]
\ .
\ee
The last term causes a shift of the effective mass to
\be \label{mshift}
m_{eff}^2 = m_0^2 + b e^{2\phi} G^{\theta\theta} N^2 = m_0^2 +
b e^{\phi- 2 B} N^2 = -2 + {2b\over 3}
\ .
\ee

The masses of the tachyons with one unit of momentum and one unit
of winding are:
\be
m_{1,0}^2 =  m_{eff}^2 + { 1 \over R^2_{S^1} } ~,~~~~~~~~
m_{0,1}^2 = m_{eff}^2 + { R_{S^1}^2 }
\ee
We see that, if $b\geq 2$, then $m_{eff}^2\geq -2/3$ and the
B-F bound is satisfied for any $R_{S^1}$. In this case the stability imposes no
restriction on $N_f/N$.

A different situation emerges for $b<2$. Consider, for example,
$b=1$. Then the tachyons
satisfy the Breitenlohner-Freedman bounds as long as
\be { 1 \over R^2_{S^1} }\geq {2\over 3}\ ,\qquad
R_{S^1}^2 \geq {2\over 3}\ ,
\ee
where the first comes from the momentum, and the second from the winding
mode.
Thus, the space is stable in the range
\be
   { 3 \over 2} \geq R_{S^1}^2 \geq { 2 \over 3}
\ ,\ee
which translates into
\be
1 \leq {N\over N_f} \leq {3\over 2}
\ .
\ee
This results in a restriction on the gauge theory that is reminiscent
of the conformal window of \cite{Seiberg},\footnote{
The Seiberg window \cite{Seiberg}
is $ {1 \over 3} < { N_c \over N_f} < { 2\over 3} $.}
%We suggest, therefore, that the stability of the $AdS_5\times S^1$ background
%will select the conformal window theories.
%While the argument for $b<2$ illustrates our point,
but we actually do not expect
the closed string tachyons to create an instability, i.e. we expect that
$b\geq 2$.

The conformal window restriction
%on $N/N_f$
should come from considering the open string tachyons on the D5-branes
because they are expected to be dual to meson operators $Q^i\tilde Q_j $,
where $i,j$ are flavor indexes.
Each meson superfield contains two scalar operators: the squark bilinear
of dimension $\Delta_1=3{N_f-N_c\over N_f}$,
and the quark bilinear of dimension
$\Delta_2=1+ 3{N_f-N_c\over N_f}$.
At the lower edge of the conformal window, $N_f=3{N_c\over 2}$,
$\Delta_1=1$,
saturating the unitarity bound \cite{Seiberg}. Within the conformal
window, ${3N_c\over 2}< N_f< 3N_c$, we find
$1< \Delta_1 <2$ and
$2 <\Delta_2 <3$.
%At
%the upper edge, $N_f= 3 N_c$, we instead find operators of dimension
%$2$ and $3$.
%analogous statements apply to the magnetic duals).
%An operator of dimension $1$
%may correspond to a scalar field in $AdS_5$ with $m^2 R_{AdS}^2=-3$,
%provided we choose the $\Delta_-$ branch for its dimension
%\cite{KW}. This value of $m^2$ does not saturate the stability bound.
%Note, however, that at the lower edge of
%the conformal window, the quark bilinear has dimension $2$
%which corresponds to $m^2 R_{AdS}^2=-4$, saturating the
%stability bound.

Now, let us consider the dual $SU(N_f-N_c)$ magnetic theory
which contains $N_f$ flavors $q^i$, $\tilde q_j$, as well as
gauge singlet free superfields $M_i^j$, with superpotential
$M_i^j q^i\tilde q_j$. In the UV theory
the superfields $q^i\tilde q_j$ contain scalar operators
of dimensions $3{ N_c\over N_f}$ and $1+ 3{ N_c\over N_f}$.
To flow to the IR fixed point we have to integrate over the superfields
$M_i^j$. This introduces a Legendre transform which changes the dimension
of an operator from $\Delta$ to $4-\Delta$ \cite{KW}.
Therefore, in the infrared theory we find scalar operators
of dimension $1+ 3{N_f-N_c\over N_f}$ and $3{N_f-N_c\over N_f}$,
in complete agreement with what was found from the electric point
of view.
%At the upper edge of the conformal window, $N_f= 3 N_c$, we find scalar
%operators of dimensions $1$ and $2$, just like at the lower edge.

Now let us see what these well-known field theory results \cite{Seiberg}
may imply about the dual string theory in AdS space. The AdS/CFT
correspondence associates a scalar operator of dimension
$\Delta$ with a field in $AdS_5$ of
$ (m R_{AdS})^2=\Delta(\Delta-4)$. Substituting the
dimensions $\Delta_1,\Delta_2$ we find
\begin{equation}\label{masses}
(m_1 R_{AdS})^2 = \left ( 3{N_c\over N_f}-1\right )^2 -4
\ ,\qquad\qquad
(m_2 R_{AdS})^2 = \left (2- 3{N_c\over N_f}\right )^2 -4
\ .
\end{equation}
These formulae have an interesting structure.
We observe that the Seiberg duality
transformation $N_c\rightarrow N_f-N_c$
interchanges $(m_1 R_{AdS})^2$ and $(m_2 R_{AdS})^2$.
Since $\Delta_1$ lies in the ``negative branch''
range \cite{KW} (between $1$ and $2$), while
$\Delta_2$ lies in the ``positive branch''
range (between $2$ and $3$),\footnote{These properties of
the meson operator dimensions were 
pointed out in \cite{Schnitzer}.} 
the Seiberg duality in effect changes
the quantization conditions for each field:
%Thus the action of the Seiberg duality in the dual AdS theory
%is the interchange of the open string tachyons of mass $m_1$
%and $m_2$ accompanied by
$\Delta_i\to 4-\Delta_i$, $i=1,2$.\footnote{ For $N_f=2 N_c$,
where the Seiberg duality maps the gauge theory into itself, we
find $ (m_1 R_{AdS})^2 = (m_2 R_{AdS})^2 =-15/4$. One of these
fields corresponds to an operator of dimension $\Delta_-=3/2$,
while the other to $\Delta_+=5/2$. Two scalar fields with $(m
R_{AdS})^2 =-15/4$ are found in the spectrum of IIB supergravity
on $AdS_5\times T^{11}$, which is dual to the $SU(N)\times SU(N)$
gauge theory on D3-branes at the tip of the conifold \cite{KW}.
This gauge theory is also mapped into itself by the Seiberg
duality, and the $SU(N)$ theory with $2 N$ flavors is obtained
when we take the coupling of one of the gauge groups to zero.
In this limit the radius of $AdS_5\times T^{11}$ becomes small,
and the supergravity approximation breaks down. }
For this operation to be admissible, it is necessary that $m_1^2,
m_2^2$ lie in the range where both the $\Delta_+$ and the
$\Delta_-$ quantizations ($\Delta_\pm= 2\pm \sqrt{4+ (m
R_{AdS})^2}$) are allowed \cite{BF,KW}:
\begin{equation} \label{range}
-4 < (m R_{AdS})^2 < -3
\ .
\end{equation}
Outside of this range $\Delta_-$ violates the unitarity bound.
Note also that at the edges of the conformal window one
of the two masses saturates the BF bound.
Imposing the condition (\ref{range}) on (\ref{masses}) we indeed find
that $N_f/N_c$ is restricted to the conformal window.
We propose therefore that, if the complete string dual
of the $SU(N_c)$ superconformal gauge theories with
$N_f$ flavors is constructed,
then the condition (\ref{range}) on the open string tachyons
will restrict it to the conformal window.

The specifics of this construction
remain to be worked out, of course.
In particular, it is not clear why the open string tachyon, whose bare
$m^2=-{1\over 2 \alpha'}$, i.e. $m^2 R_{AdS}^2=-3$, can approach the
BF bound. Perhaps the $m^2$ receives a negative shift due to interaction
with the R-R fluxes or $\alpha'$ corrections. In any case, it is clear
that the 2-derivative effective action is mainly a qualitative tool,
and that a more precise treatment of D-branes on the cigar is necessary.
Another important difficulty is the fact that
there is a large number of flavor branes, which can lead to strong
coupling effects for open strings in the bulk. In other words, we
find that $ e^{\phi} N_f \sim {2 \over 3} $, which is not small.

Finally, we would like to speculate that T-duality on the circle $S^1$ is
related to Seiberg's electric-magnetic  duality \cite{Seiberg}.
%Finally, we would like to speculate that
%Seiberg's electric-magnetic  duality \cite{Seiberg} is realized as
%the T-duality on the circle $S^1$ accompanied by the change
%of the quantization conditions in $AdS_5$ for the open string
%tachyons, $\Delta\rightarrow 4-\Delta$.
This is suggested by
eq. (\ref{masses}) and the discussion below it. It is tempting
to think that $m_1^2$ and $m_2^2$ refer to states of the open string tachyon
with some momentum or some winding. Then it is natural that such states are
interchanged under the T-duality.\footnote{For this it is necessary
that the branes introducing flavors are themselves invariant under the
T-duality. Therefore, the correct string construction
should probably include D4-branes and D5-branes.}
In the 2-derivative effective action approximation, after the T-duality
we have a solution of the same form as the one we started with but with
$N_f \to \sqrt{ 2 \over 3} N$ and $N \to \sqrt{ 3 \over 2} N_f$.
This transformation preserves both the radius of the $AdS_5$ and the effective
5-d coupling. However, it
 is not of the same form as the Seiberg duality \cite{Seiberg} because
the latter interchanges $N_c$ with $N_f-N_c$.
It is amusing to observe that, if we denote $N_c = \sqrt{ 2 \over 3} N$,
and replace $N_f$ by $N_f - \sqrt{ 2 \over 3} N$
throughout our paper, then the T-duality
transformation becomes identical to the Seiberg duality.
We hope that
a better analysis of D-branes on the cigar could
explain the required shift of $N_f$.
%%%%%%%%%%%%%%%%%%%%%%%%%%%%%%%%%%%%%%%%%%%%%%%
\section*{Acknowledgments}
%%%%%%%%%%%%%%%%%%%%%%%%%%%%%%%%%%%%%%%%%%%%%%%%%%%%
We are grateful to the Aspen Center of Physics where this
project was started in 2000.
We also thank A.M. Polyakov and N. Seiberg for useful
discussions, and S. Kuperstein and J. Sonnenschein for an
important communication after the original version of this paper
was submitted. 
This material is based upon work
supported by the National Science Foundation Grants No.
PHY-0243680 and PHY-0140311.
We acknowledge support from the U.S.\ Department of Energy under
grant  DE-FG02-90ER40542.
Any opinions, findings, and conclusions or recommendations expressed in
this material are those of the authors and do not necessarily reflect
the views of the National Science Foundation.

\begingroup\raggedright\endgroup
\end{document}